\documentclass[prl,twocolumn,preprintnumbers,superscriptaddress,amsmath,amssymb]{revtex4-1}
\usepackage{graphicx}
\usepackage{subfigure}
\usepackage{mathrsfs}
\usepackage{amsfonts}
\usepackage{times}
\usepackage{amsmath}
\usepackage{leftidx}
\usepackage{tikz}
\usepackage{color}
\usepackage[colorlinks,linkcolor=blue,citecolor=blue]{hyperref}

\newcommand{\Tr}{\operatorname{Tr}}

\usepackage{bbold}
\usepackage{braket}
\usepackage{mathtools}

\begin{document}
\title{Universal Error Bound for Constrained Quantum Dynamics}
\author{Zongping Gong}
\affiliation{Department of Physics, University of Tokyo, 7-3-1 Hongo, Bunkyo-ku, Tokyo 113-0033, Japan}
\author{Nobuyuki Yoshioka}
\affiliation{Department of Physics, University of Tokyo, 7-3-1 Hongo, Bunkyo-ku, Tokyo 113-0033, Japan}
\affiliation{Theoretical Quantum Physics Laboratory, RIKEN Cluster for Pioneering Reserach (CPR), Wako-shi, Saitama 351-0198, Japan}
\author{Naoyuki Shibata}
\affiliation{Department of Physics, University of Tokyo, 7-3-1 Hongo, Bunkyo-ku, Tokyo 113-0033, Japan}
\author{Ryusuke Hamazaki}
\affiliation{Department of Physics, University of Tokyo, 7-3-1 Hongo, Bunkyo-ku, Tokyo 113-0033, Japan}
\affiliation{Nonequilibrium Quantum Statistical Mechanics RIKEN Hakubi Research Team, RIKEN Cluster for Pioneering Research (CPR), RIKEN iTHEMS, Wako, Saitama 351-0198, Japan}
\date{\today}

\begin{abstract}
It is well known in quantum mechanics that a large energy gap between a Hilbert subspace of specific interest and the remainder of the spectrum can suppress transitions from the quantum states inside the subspace to those outside due to additional couplings that mix these states, and thus approximately lead to a constrained dynamics within the subspace. While this statement has widely been used to approximate quantum dynamics in various contexts, a general and quantitative justification stays lacking.  Here we establish an observable-based error bound for such a constrained-dynamics approximation in generic gapped quantum systems. This universal bound is a linear function of time that only involves the energy gap and coupling strength, provided that the latter is much smaller than the former. We demonstrate that either the intercept or the slope in the bound is asymptotically saturable by simple models. We generalize the result to quantum many-body systems with local interactions, for which the coupling strength diverges in the thermodynamic limit while the error is found to grow no faster than a power law $t^{d+1}$ in $d$ dimensions. Our work establishes a universal and rigorous result concerning nonequilibrium quantum dynamics.
\end{abstract}
\maketitle

\emph{Introduction.---} Approximations appear ubiquitously in science and their validity should be justified by error estimations \cite{Cheney1966}. In quantum physics, where only a few systems are exactly solvable \cite{Albeverio1988}, one of the most widely used approximations is based on the separation of energy scales or the existence of large energy gaps \cite{Sakurai2011}. While an entire quantum system can be very complicated, it can dramatically be simplified by keeping only the degrees of freedom with relevant energy scales. Technically, this is achieved by projecting the full Hamiltonian onto a Hilbert subspace. Two prototypical examples are approximating atoms as few-level systems in quantum optics \cite{Scully1997} and crystalline materials as few-band systems in condensed matter physics \cite{Ashcroft1976}. When we add a weak coupling term such as external fields or interactions, it suffices to consider the action restricted in the subspace, provided that the manifold is  energetically well-isolated from the remaining. It is well-known from the perturbation theory that the error is of the order of the inverse energy gap, and thus vanishes in the infinite gap limit \cite{Kato1966}. 

The suppression of error by energy gap has tacitly been used for approximating not only static quantum states but also quantum dynamics \cite{Polkovnikov2011}. For example, quench dynamics in the Bose- and Fermi-Hubbard models are implemented by ultracold atoms in deep optical lattices such that the projection onto the ground-state band is a good approximation \cite{Jaksch1998,Hofstetter2002,Bloch2012b,Gross2020}. More recently, the peculiarly slow thermalization dynamics observed in a strongly interacting Rydberg-atom chain is actively studied on the basis of the so-called PXP model, where the constrained dynamics is within the Hilbert subspace with adjacent Rydberg excitations forbidden \cite{Bernien2017,Turner2018}. 

While the approximation of constrained dynamics from large energy gaps has widely been used in the literature, a general and quantitative justification stays lacking. Here, we fill this gap by deriving a universal error bound for constrained dynamics. This bound is simply a linear function of time and depends only on the coupling strength and energy gap. Our main strategy is a general perturbative analysis based on the \emph{Schrieffer-Wolff transformation} (SWT) \cite{Schrieffer1966}, which is a unitary transformation that block diagonalizes the perturbed Hamiltonian and has been used to estimate the errors in equilibrium setups \cite{Bravyi2011}. We mostly focus on bounded coupling terms, but will also outline the many-body generalization based on locality. In addition to the Lieb-Robinson bound \cite{Lieb1972,Nachtergaele2006,Bravyi2006}, the quantum speed limit \cite{Margolus1998,Taddei2013,delCampo2013,Deffner2013}, the bound on energy absorption for Floquet systems \cite{Abanin2015b,Mori2016} and the bound on chaos \cite{Stanford2016}, our work contributes yet another rigorous and universal bound in nonequilibrium quantum dynamics.

\begin{figure}
\begin{center}
\includegraphics[width=8.5cm]{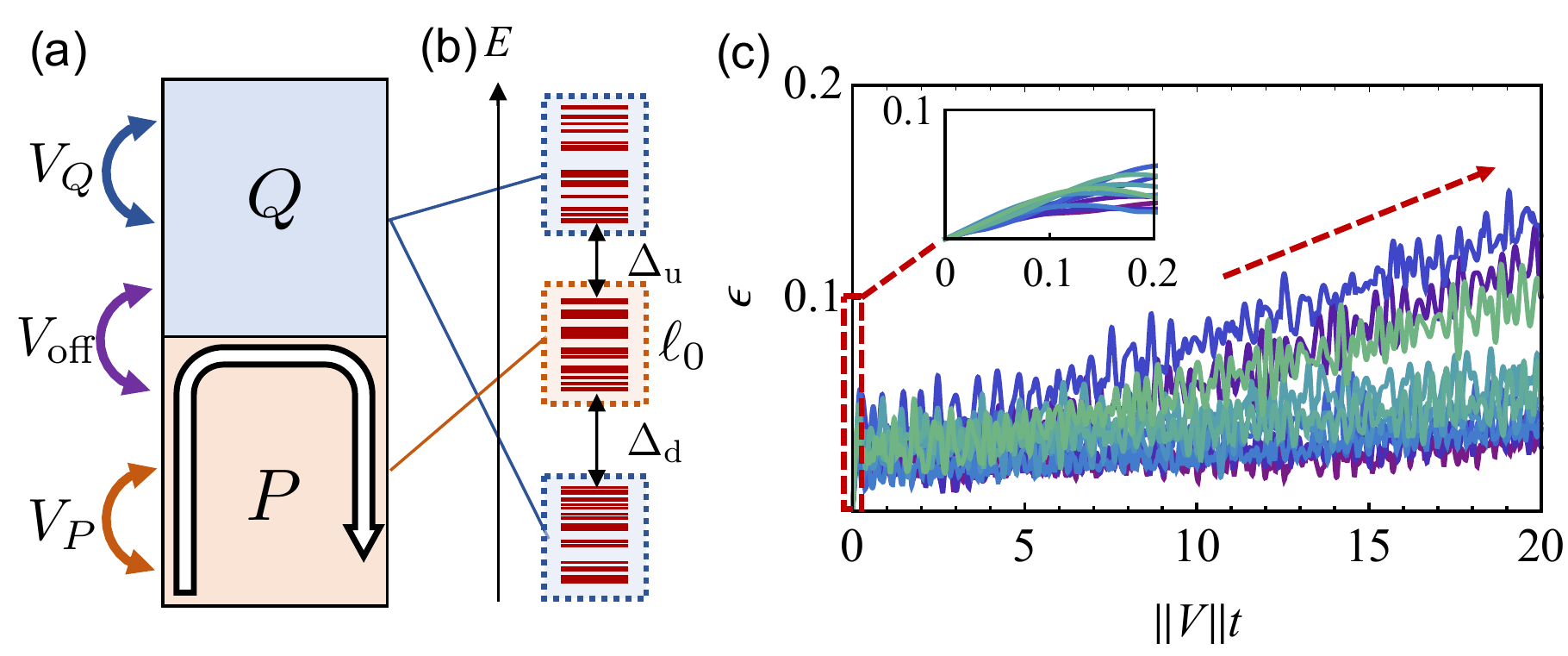}
\end{center}
\caption{(a) Schematic illustration of the setup. The original Hamiltonian $H_0$ has an isolated energy band onto which the projector is $P$, whose complement is $Q=1-P$. A general coupling can be decomposed into $V=V_P+V_Q+V_{\rm off}$, where $V_P\equiv PVP$, $V_Q\equiv QVQ$ and $V_{\rm off}\equiv PVQ+QVP$. The arrow within $P$ refers to a constrained dynamics. (b) Typical energy spectrum of $H_0$ with an isolated band $\ell_0$. The energy gap is defined as $\Delta_0\equiv\min\{\Delta_{\rm u},\Delta_{\rm d}\}$. (c) Error dynamics (\ref{et}) in a random model with $\Delta_0=10\|V\|$. Different colors correspond to different realizations. The rectangle and the arrow indicate the initial sudden jump and the subsequent linear growth, respectively. Inset: Zoom in of the initial jump.}
\label{fig1}
\end{figure}

\emph{Setup and numerical trials.---} We consider a quantum system with arbitrarily large Hilbert-space dimension described by $H_0$, which has an isolated energy band $\ell_0$ separated from the remainder of the spectrum by an energy gap $\Delta_0$ (see Fig.~\ref{fig1}(b)). The projector onto the subspace spanned by all the eigenstates in $\ell_0$ is denoted as $P$ (see Fig.~\ref{fig1}(a)). With an additional perturbation term $V$ added to $H_0$, the total Hamiltonian becomes
\begin{equation}
H=H_0+V.
\label{H0V}
\end{equation}
To quantify the deviation between the constrained dynamics generated by the projected Hamiltonian $H_P\equiv PHP$ and the actual dynamics starting from a state in $\ell_0$, we define the error with respect to an observable $O$ as 
\begin{equation}
\epsilon(t)\equiv\|P(e^{iHt}Oe^{-iHt}-e^{iH_Pt}Oe^{-iH_Pt})P\|,
\label{et}
\end{equation}
where $\|\cdot\|$ denotes the operator norm, i.e., the largest singular value. Without loss of generality, we assume $O$ to be normalized as $\|O\|=1$. While only Hermitian observables are directly measurable in experiments, our result applies equally to non-Hermitian operators. The maximal value of $\epsilon(t)$ over $O$ is actually the superoperator norm $\|\mathscr{P}(e^{it{\rm ad}_H}-e^{it{\rm ad}_{H_P}})\|_{\infty\to\infty}$ induced by the operator norm \cite{Wolf2015}, where $\|\mathscr{L}\|_{\infty\to\infty}\equiv\max_{\|O\|=1}\|\mathscr{L}O\|$, $\mathscr{P}O\equiv POP$ and ${\rm ad}_AO\equiv [A,O]$.

It is helpful to first form an intuition on the typical behavior of $\epsilon(t)$. We carry out numerical simulations for a randomly constructed system with three bands, each containing four levels and separated from the others by a large gap. The observable is also taken to be random, and $\ell_0$ is chosen to be the middle band. As shown in Fig.~\ref{fig1}(c), while the local fluctuations in $\epsilon(t)$ differ significantly for different random realizations of the system, there seems to be two universal features. First, $\epsilon(t)$ initially undergoes a sudden ``jump". Precisely speaking, the ``jump" is a rapid growth within an $\mathcal{O}(\Delta_0^{-1})$ time interval (see the inset in Fig.~\ref{fig1}(c)). Second, $\epsilon(t)$ grows linearly despite of irregular fluctuations. It is thus natural to conjecture that $\epsilon(t)$ is bounded by a linear function of time.

\emph{Main result and its qualitative explanation.---} The above conjecture turns out to be indeed true. In the regime $\Delta_0\gg\|V\|$ and for an intermediately long time $t\ll\frac{\Delta_0}{\|V\|^2}$ (i.e., $\|V\|t$ is considered as order one), we claim the following universal asymptotical bound:
\begin{equation}
\epsilon(t)\lesssim\frac{4\|V\|}{\Delta_0}+\frac{2\|V\|^2}{\Delta_0}t, 
\label{asymb}
\end{equation}
where ``$\lesssim$" means that there could be a tiny violation up to $\mathcal{O}(\frac{\|V\|^2}{\Delta_0^2})$. It is possible to derive a bound valid even when $\Delta_0$ is comparable with $\|V\|$, but the form is a bit involved and exactly reproduces Eq.~(\ref{asymb}) in the large-gap regime \cite{Gong2020}.

Two more remarks on the applicability of Eq.~(\ref{asymb}) are in order. First, the energy band can be embedded anywhere in the energy spectrum of $H_0$. It may consist of the ground states, mid-gap states or even the most excited states. All these situations will later be exemplified. Second, we do not assume any constraint on the width of the energy band. It can be zero (e.g., for ground-state manifolds) or even larger than $\Delta_0$.

It is rather easy to understand the orders of the intercept and the slope in Eq.~(\ref{asymb}). According to the standard perturbation theory \cite{Sakurai2011}, a state in $\ell_0$ should basically lie in $\ell$, the perturbed energy band in $H$, but also slightly contain some components outside $\ell$. These components have $\mathcal{O}(\frac{\|V\|}{\Delta_0})$ weights (amplitudes), and their rapid oscillations owing to the dynamical phases $e^{-it\mathcal{O}(\Delta_0)}$ lead to the initial ``jump" of $\epsilon(t)$. Also, the effective Hamiltonian in the Green's function (Fourier transform) of the projected unitary $Pe^{-iHt}P$ is known to be $H_{\rm eff}(\omega)=H_P+\Sigma(\omega)$, where the self-energy $\Sigma(\omega)=PVQ(\omega-H_Q)^{-1}QVP$ is of the order of $\frac{\|V\|^2}{\Delta_0}$ for $\omega\in\ell_0$ \cite{Feshbach1958,Feshbach1962,Brion2007}.  However, it is unclear from the above argument why the factors before $\frac{\|V\|}{\Delta_0}$ and $\frac{\|V\|^2}{\Delta_0}$ should be $4$ and $2$. It is even unclear why these factors can be finite, even though there can be a very large number of levels in or/and outside $\ell_0$.

\emph{``Worst" models.---} Before deriving the main result, let us comment on the tightness of the bound (\ref{asymb}). We emphasize that the bound is \emph{universally} valid, so the tightness should be analyzed for the ``\emph{worst}" models and observables instead of the typical ones like the random model in Fig.~\ref{fig1}. It turns out that, \emph{separately}, both of the constant and the time-linear terms are tight and can asymptotically (in the large gap limit) be saturated in very simple models.

\begin{figure}
\begin{center}
\includegraphics[width=7cm]{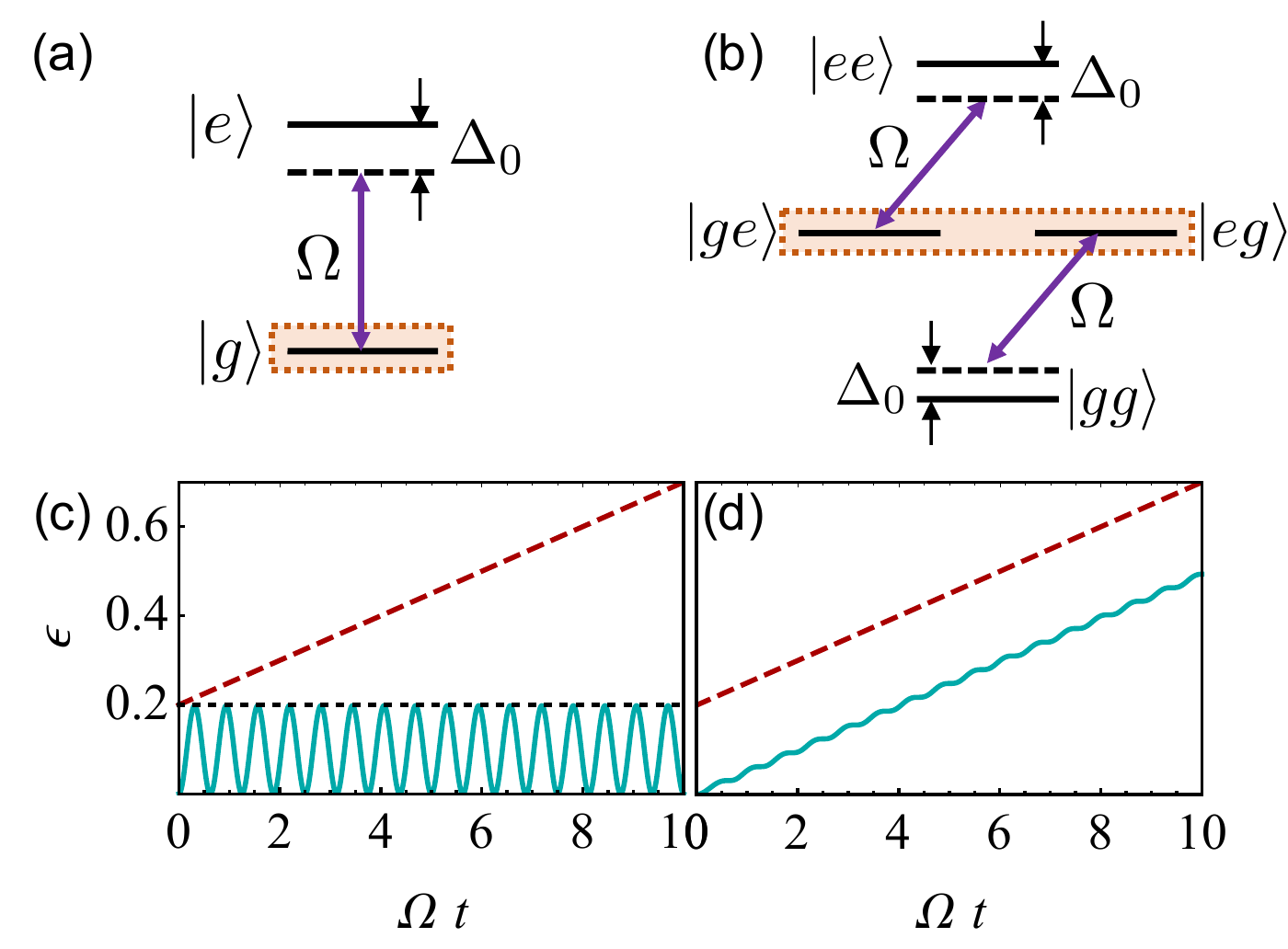}
\end{center}
\caption{``Worst" models which realize the saturation of (a) the intercept and (b) the slope in Eq.~(\ref{asymb}). (c) and (d) are the corresponding error dynamics (green solid curves) of (a) and (b) given in Eqs.~(\ref{wst1}) and (\ref{wst2}) with $\Delta_0=10\Omega$. The black dotted line indicates the intercept and the red dashed lines are the asymptotic bound (\ref{asymb}).}
\label{fig2}
\end{figure}

We first demonstrate the saturation of the constant in a two-level atom driven by a classical laser with detuning $\Delta_0$ and Rabi frequency $\Omega$ (see Fig.~\ref{fig2}(a)). In this case, $H_0=\frac{\Delta_0}{2}\sigma^z$ and $V=\frac{\Omega}{2}\sigma^x$ \cite{RF} so that $\|V\|=\frac{\Omega}{2}$, where $\sigma^x=|e\rangle\langle g|+|g\rangle\langle e|$ and $\sigma^z=|e\rangle\langle e|-|g\rangle\langle g|$ are the Pauli matrices. Choosing $P=\frac{1}{2}(1-\sigma^z)=|g\rangle\langle g|$ to be the ground-state projector and $O=\sigma^x$, we can easily calculate the error to be
\begin{equation}
\epsilon(t)=\frac{\Delta_0\Omega}{\Delta^2}|1-\cos(\Delta t)|,
\label{wst1}
\end{equation}
where $\Delta=\sqrt{\Delta_0^2+\Omega^2}$. This quantity rapidly reaches its maximum $\frac{2\Omega\Delta_0}{\Delta^2}$ at $t=\frac{\pi}{\Delta}$, which asymptotically saturates $\frac{4\|V\|}{\Delta_0}=\frac{2\Omega}{\Delta_0}$ in the large $\Delta_0$ limit. The error dynamics stays exactly the same for $P=\frac{1}{2}(1+\sigma^z)=|e\rangle\langle e|$.

We move on to demonstrate the saturation of the slope. Here we should consider a situation with $\Tr P>1$, otherwise $\epsilon(t)$ never exceeds an $\mathcal{O}(\frac{\|V\|}{\Delta_0})$ constant even in the long-time limit \cite{Gong2020}. It turns out that a four-level system with $H_0=\frac{\Delta_0}{2}(\sigma^z_1+\sigma^z_2)$ and $V=\frac{\Omega}{2}\sigma^x_1$, which describes two identical two-level atoms with only one driven by a classical laser (see Fig.~\ref{fig2}(b)), already constitutes such a worst-case scenario. Choosing $P=\frac{1}{2}(1-\sigma^z_1\sigma^z_2)$ (with $\Tr P=2$) and $O=\frac{1}{2}(\sigma^x_1\sigma^x_2+\sigma^y_1\sigma^y_2)$ ($\sigma^y\equiv i(|g\rangle\langle e|-|e\rangle\langle g|)$), we have
\begin{equation}
\epsilon(t)=\left|\left[\cos\left(\frac{\Delta t}{2}\right)+i\frac{\Delta_0}{\Delta}\sin\left(\frac{\Delta t}{2}\right)\right]^2-e^{i\Delta_0 t}\right|,
\label{wst2}
\end{equation}
which is well approximated by $|e^{i(\Delta-\Delta_0)t}-1|\simeq\frac{\Omega^2}{2\Delta_0}t=\frac{2\|V\|^2}{\Delta_0}t$ for $t\ll\frac{\Delta_0}{\|V\|^2}=\frac{4\Delta_0}{\Omega^2}$. We plot Eqs.~(\ref{wst1}) and (\ref{wst2}) in Figs.~\ref{fig2}(c) and (d), respectively, where the asymptotic bound (\ref{asymb}) is also shown for comparison.

\emph{Derivation of the main result.---} We turn to the derivation of Eq.~(\ref{asymb}). The first step is to perform the SWT \cite{Schrieffer1966,Bravyi2011}: 
\begin{equation}
SHS^\dag= H_0+V_{\rm diag}+V',
\label{SWT}
\end{equation}
where $S=e^T$ is unitary and the anti-Hermitian generator $T$ is determined from ${\rm ad}_T(H_0+V_{\rm diag})=-V_{\rm off}$, with $V_{\rm diag}\equiv PVP+QVQ$ and $V_{\rm off}\equiv V-V_{\rm diag}$ being the block-diagonal and block-off-diagonal components of $V$, respectively. Since $H_0$ is block diagonalized while $V_{\rm off}$ is off-block diagonalized, it follows that $T$ can be restricted to be off-block diagonalized to satisfy the following \emph{Sylvester equation} \cite{Sylvester1884}: 
\begin{equation}
T H_Q-H_PT=-PVQ.
\end{equation}
Provided that the spectra of $H_P$ and $H_Q$ are separated by a gap $\Delta$, $\|T\|$ is rigorously upper bounded by \cite{Bravyi2011,Bhatia1997}
\begin{equation}
\|T\|\le\frac{\|PVQ\|}{\Delta}\lesssim\frac{\|V\|}{\Delta_0}.
\label{TVD}
\end{equation}
We recall that ``$\lesssim$" in Eq.~(\ref{TVD}) allows a tiny violation with $\mathcal{O}(\frac{\|V\|^2}{\Delta_0^2})$, and is validated by $\Delta\ge\Delta_0-2\|V\|$, a result ensured by \emph{Weyl's perturbation theorem} \cite{Weyl1912}. Accordingly, the norm of the remaining term 
\begin{equation}
V'=\sum^\infty_{n=1}\frac{n}{(n+1)!}{\rm ad}_T^nV_{\rm off}
\end{equation}
in Eq.~(\ref{SWT}) should asymptotically be bounded by 
\begin{equation}
\|V'\|\lesssim\frac{1}{2}\|[T,V_{\rm off}]\|\le\|T\|\|V\|\lesssim\frac{\|V\|^2}{\Delta_0} 
\label{Vpb}
\end{equation}
in the large gap regime. Here we have used the fact that $\|T\|$ is small and $\|V_{\rm off}\|\le\|V\|$ \cite{Gong2020}.

With the help of the SWT, we can rewrite the error (\ref{et}) as
\begin{equation}
\epsilon(t)=\|P[S_{H_1}(t)^\dag L(t)SOS^\dag L(t)^\dag S_{H_1}(t)-O]P\|,
\label{rw}
\end{equation}
where $H_1\equiv H_0+V_{\rm diag}$, $L(t)=e^{-iH_1t}e^{i(H_1+V')t}$ is the Loschmidt-echo operator \cite{Prosen2006} and $S_{H_1}(t)=e^{-iH_1t}e^Te^{iH_1t}=e^{e^{-iH_1t}Te^{iH_1t}}$ is the SWT in the interacting picture. This rewritten form (\ref{rw}) has a crucial property that the generators of $S$ and $S_{H_1}(t)$ both have small norms (at most) of the order of $\frac{\|V\|}{\Delta_0}$, and so is $L(t)=\overrightarrow{{\rm T}}e^{i\int_0^tdt'e^{-iH_1t'}V'e^{iH_1t'}}$ ($\overrightarrow{{\rm T}}$: time ordering) \cite{dLt} for a time scale of interest (i.e., $\|V\|t$ is of order one). Applying the inequality $\|(\prod_\alpha U_\alpha)O(\prod_\alpha U_\alpha)^\dag-O\|\le\sum_\alpha\|U_\alpha OU_\alpha^\dag-O\|$ for unitaries $U_\alpha$'s \cite{Gong2020} to Eq.~(\ref{rw}), we obtain
\begin{equation}
\begin{split}
\epsilon(t)\le\|SOS^\dag-O\|&+\|L(t)OL(t)^\dag-O\| \\
&+\|S_{H_1}(t)^\dag OS_{H_1}(t)-O\|.
\end{split}
\label{dec}
\end{equation}
Using the inequality \cite{Gong2020}
\begin{equation}
\|e^{T}Oe^{-T}-O\|\le\|[T,O]\|\le 2\|T\| 
\label{TOT}
\end{equation}
for all $T=-T^\dag$ (and $\|O\|=1$), we can bound the first and the third terms in Eq.~(\ref{dec}) by $2\|T\|$, and the second term by
\begin{equation}
\begin{split}
\|L(t)OL(t)^\dag-O\|&\le\int^t_0dt'\|[e^{-iH_1t'}V'e^{iH_1t'},O]\| \\
&\le2\|V'\|t.
\end{split}
\label{LOL}
\end{equation}
Substituting these exact bounds in Eqs.~(\ref{TOT}) and (\ref{LOL}) and those asymptotic ones in Eqs.~(\ref{TVD}) and (\ref{Vpb}) into Eq.~(\ref{dec}), we obtain Eq.~(\ref{asymb}). Now it is clear that the constant in Eq.~(\ref{asymb}) arises from the SWT and the time-evolved SWT, i.e., the first and third terms on the rhs of Eq.~(\ref{dec}), while the time-linear term arises from the Loschmidt echo, i.e., the second term on the rhs of Eq.~(\ref{dec}). 

\emph{Generalization to many-body systems.---} We next focus on quantum many-body systems defined on a general $d$-dimensional lattice $\Lambda$, where each site is associated with a finite dimensional local Hilbert space.  A particularly important case is \emph{locally} interacting systems, whose many-body Hamiltonians still take the form of Eq.~(\ref{H0V}) while both $H_0$ and $V$ are a sum of Hermitian operators supported on finite regions, whose norms are uniformly bounded. Formally, we can write $V=\sum_{A\subseteq\Lambda}V_A$ with $V_A$ supported on a connected region $A$ and $V_A=0$ if its volume $|A|$ exceeds a threshold. In contrast with the few-body systems, the error bound on the rhs of Eq.~(\ref{asymb}) diverges in the thermodynamic limit. This is because $\|V\|$ grows linearly with respect to the system volume $|\Lambda|\sim L^d$, where $L=l_\Lambda$ is the diameter of the entire system.

However, by further assuming (i) $H_0=\sum_j H_{0j}$ is commuting and frustration-free in the sense that all the local operators commute, i.e., $[H_{0j},H_{0j'}]=0$ for all $j,j'\in\Lambda$, and all the global ground states minimize local energies everywhere; (ii) $O=O_X$ is a local observable supported on $X$ with $|X|,l_X\sim\mathcal{O}(1)$, we can still derive a meaningful bound: 
\begin{equation}
\epsilon(t)\le \frac{\|V\|_{\star}}{\Delta_0}p(t),
\label{mbb}
\end{equation}
where $\|V\|_{\star}\equiv\max_{j\in\Lambda}\sum_{A\ni j}\|V_A\|$ is the \emph{local} interaction strength, which is set to be $\mathcal{O}(1)$ by rescaling the time scale, and $p(t)$ is a polynomial of $t$ with degree $d+1$ and (at most) order-one coefficients. This result implies that for a prescribed precision $\epsilon$, the constrained dynamics is a \emph{locally} good approximation up to a time scale (at least) of $\mathcal{O}((\Delta_0\epsilon)^{\frac{1}{d+1}})$.

Let us outline the proof of Eq.~(\ref{mbb}), whose full details are available in Ref.~\cite{Gong2020}. The general idea is to combine the \emph{local} SWT \cite{Bravyi2011,Datta1996} and the Lieb-Robinson bound \cite{Lieb1972}. By local we mean that the generator $T$ is a sum of local operators. The locality of $T$ and $O_X$ allows us to modify Eq.~(\ref{TOT}) into \cite{Gong2020}
\begin{equation}
\|e^TO_Xe^{-T}-O_X\|\le2|X|\|T\|_\star.
\end{equation}
Similar to Eq.~(\ref{TVD}), $\|T\|_\star$ can be upper bounded by $\mathcal{O}(\frac{\|V\|_\star}{\Delta_0})$, and so can the first term on the rhs of Eq.~(\ref{dec}). As for the third term, we note that $\|S_{H_1}(t)^\dag O_XS_{H_1}(t)-O_X\|=\|e^{-T}O_X^{H_1}(t)e^T-O_X^{H_1}(t)\|$ with $O_X^{H_1}(t)\equiv e^{iH_1t}O_Xe^{-iH_1t}$ being the observable in the Heisenberg picture. While the support of $O_X^{H_1}(t)$ generally covers the entire lattice in a rigorous sense, we can show from the Lieb-Robinson bound \cite{Lieb1972} that the support volume is effectively of the order of $(l_X+2vt)^d$ \cite{Gong2020}, where $v$ is the Lieb-Robinson velocity. We emphasize that $v$ is essentially determined by $V$ since $H_0$ is by assumption commuting, i.e., consists of commutative local operators, and thus almost does not contribute to the spreading of operators \cite{esal}. This fact ensures the finiteness of $v$ even in the infinite-gap limit, where the usual Lieb-Robinson velocity \cite{Lieb1972} determined from $H$ diverges.

Moreover, the locality of $T$ in turn ensures the (quasi-)locality of $V'$ in Eq.~(\ref{SWT}), and we can show that $\|V'\|_\star$ is no more than $\mathcal{O}(\frac{\|V\|_\star^2}{\Delta_0})$ in the large gap regime, just like Eq.~(\ref{Vpb}). Following the same argument used for bounding $\|S_{H_1}(t)^\dag O_XS_{H_1}(t)-O_X\|$, we can show that the order of the integrand in Eq.~(\ref{LOL}) is no more than $\|V'\|_\star(l_X+2vt')^d$, whose integral is a polynomial of $t$ with degree $d+1$. Combining all the analyses above, we obtain Eq.~(\ref{mbb}) from Eq.~(\ref{dec}). 

\begin{figure}[t]
\begin{center}
\includegraphics[width=7cm]{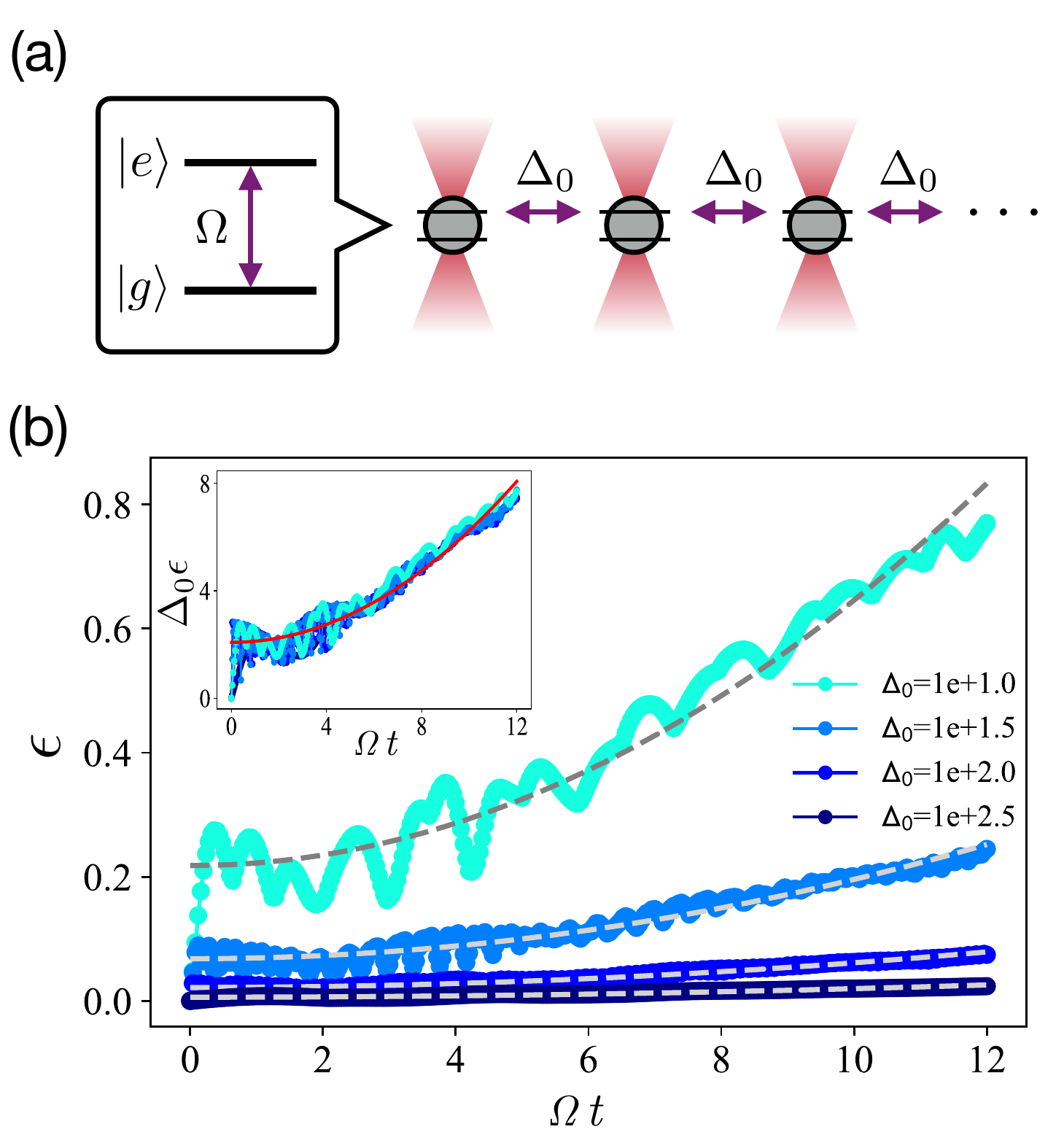}
\end{center}
\caption{(a) An array of two-level atoms with interaction $\Delta_0$ between adjacent excited states. 
Each atom is resonantly driven with an identical Rabi frequency $\Omega$. (b) The quadratic growth of $\epsilon(t)$ concerning the error dynamics of $O=\sigma^y_{j=1}$ in the parent Hamiltonian of the PXP model defined by Eq.~\eqref{eqn:pxp_def} with $\log_{10}\Delta_0=1.0, 1.5, 2.0, 2.5$. The best fitting quadratic curves (grey dotted lines) are confirmed to be more accurate than the linear ones, while the growth becomes rather linear after the correlation spreads throughout the system at $\Omega t\sim 12$. The Rabi frequency is $\Omega=2$ and the system size is $L=12$. Inset: Rescaled error $\Delta_0 \epsilon(t)$, whose quasi-independence on $\Delta_0$ is consistent with Eq.~\eqref{mbb}.}
\label{fig3}
\end{figure}

To illustrate our findings, we consider the error dynamics in the parent Hamiltonian of the PXP model~\cite{Turner2018}. As is graphically illustrated in Fig.~\ref{fig3}(a), the Hamiltonian is given as 
\begin{equation}\label{eqn:pxp_def}
H_0=\frac{\Delta_0}{4}\sum_{j=1}^{L-1}(\sigma^z_j+1)(\sigma^z_{j+1}+1),\;\;\;\;
V=\frac{\Omega}{2}\sum_j\sigma^x_j,
\end{equation}
where $L$ is the system size. The constrained dynamics concerns $P=\prod_j\left[1- \frac{1}{4}(1+\sigma_j^z)(1+\sigma_{j+1}^z)\right]$, which is a projector onto the Hilbert subspace with adjacent excitations forbidden.
As shown in Fig.~\ref{fig3}(b),  $\epsilon(t)$ indeed grows like $t^2$ rather than $t$ after the ``sudden" jump, implying that the power bound $t^{d+1}$ presented in Eq.~\eqref{mbb} is \emph{qualitatively} tight. 
On the other hand, we do not expect quantitative saturation of the many-body bound due to the looseness of the Lieb-Robinson bound \cite{Lieb1972}.
We also argue that beyond the Lieb-Robinson time $t^*\sim L/v$  the error grows linearly.
This is because the correlation spreads throughout the system and hence the bound based on the locality argument no longer holds.

\emph{Summary and outlook.---} In summary, we have established a universal and tight error bound (\ref{asymb}) for constrained dynamics in generic quantum systems with isolated energy bands. By universal we mean that the bound is generally applicable and only involves a minimal number of parameters (coupling strength and energy gap). By tight we mean that it can partially be saturated in some worst cases. The result has been generalized to many-body systems by means of the Lieb-Robinson bound. It is found that the error of a local observable grows no faster than $t^{d+1}$, so many-body constrained dynamics can stay \emph{locally} good approximations.

The error bound can readily be generalized to open quantum systems with decoherence-free subspaces \cite{Zanardi1997,Lidar1998} subject to coherent perturbations, as has been shown in Ref.~\cite{Gong2020}. The obtained result is actually related to the quantum Zeno effect \cite{Misra1977,Knight2000,Facchi2002}. Our work also raises many open questions such as whether the intercept and the slope in Eq.~(\ref{asymb}) can simultaneously be saturated and whether it is possible to generalize to open quantum many-body systems with an extensive number of dark states \cite{Zoller2014,Gong2017}. Other directions of future studies include the generalizations to higher-order SWTs \cite{Bravyi2011,Gong2020} and long-range interacting systems \cite{Hauke2013,Eisert2013,Gorshkov2014b,Lucas2019,Kuwahara2019}.


We acknowledge Takashi Mori for valuable comments.
The numerical calculations were carried out with the help of QuTiP~\cite{qutip}.
Z.G. was supported by MEXT. N.Y. and R.H.
were supported by Advanced Leading Graduate Course for Photon Science (ALPS) of Japan Society for the Promotion of Science (JSPS). N.Y. was supported by JSPS KAKENHI Grant-in-Aid for JSPS fellows Grant No. JP17J00743. N.S. acknowledges support of the Materials Education program for the future leaders in Research, Industry, and Technology (MERIT). R.H. was supported by JSPS KAKENHI Grant-in-Aid for JSPS fellows Grant No. JP17J03189).

\bibliography{GZP_references}

\begin{thebibliography}{51}%
\makeatletter
\providecommand \@ifxundefined [1]{%
 \@ifx{#1\undefined}
}%
\providecommand \@ifnum [1]{%
 \ifnum #1\expandafter \@firstoftwo
 \else \expandafter \@secondoftwo
 \fi
}%
\providecommand \@ifx [1]{%
 \ifx #1\expandafter \@firstoftwo
 \else \expandafter \@secondoftwo
 \fi
}%
\providecommand \natexlab [1]{#1}%
\providecommand \enquote  [1]{``#1''}%
\providecommand \bibnamefont  [1]{#1}%
\providecommand \bibfnamefont [1]{#1}%
\providecommand \citenamefont [1]{#1}%
\providecommand \href@noop [0]{\@secondoftwo}%
\providecommand \href [0]{\begingroup \@sanitize@url \@href}%
\providecommand \@href[1]{\@@startlink{#1}\@@href}%
\providecommand \@@href[1]{\endgroup#1\@@endlink}%
\providecommand \@sanitize@url [0]{\catcode `\\12\catcode `\$12\catcode
  `\&12\catcode `\#12\catcode `\^12\catcode `\_12\catcode `\%12\relax}%
\providecommand \@@startlink[1]{}%
\providecommand \@@endlink[0]{}%
\providecommand \url  [0]{\begingroup\@sanitize@url \@url }%
\providecommand \@url [1]{\endgroup\@href {#1}{\urlprefix }}%
\providecommand \urlprefix  [0]{URL }%
\providecommand \Eprint [0]{\href }%
\providecommand \doibase [0]{http://dx.doi.org/}%
\providecommand \selectlanguage [0]{\@gobble}%
\providecommand \bibinfo  [0]{\@secondoftwo}%
\providecommand \bibfield  [0]{\@secondoftwo}%
\providecommand \translation [1]{[#1]}%
\providecommand \BibitemOpen [0]{}%
\providecommand \bibitemStop [0]{}%
\providecommand \bibitemNoStop [0]{.\EOS\space}%
\providecommand \EOS [0]{\spacefactor3000\relax}%
\providecommand \BibitemShut  [1]{\csname bibitem#1\endcsname}%
\let\auto@bib@innerbib\@empty
\bibitem [{\citenamefont {Cheney}(1966)}]{Cheney1966}%
  \BibitemOpen
  \bibfield  {author} {\bibinfo {author} {\bibfnamefont {E.~W.}\ \bibnamefont
  {Cheney}},\ }\href@noop {} {\emph {\bibinfo {title} {Introduction to
  Approximation Theory}}}\ (\bibinfo  {publisher} {McGraw-Hill, New York},\
  \bibinfo {year} {1966})\BibitemShut {NoStop}%
\bibitem [{\citenamefont {Albeverio}\ \emph {et~al.}(1988)\citenamefont
  {Albeverio}, \citenamefont {Gesztesy}, \citenamefont {H{\o}egh-Krohn},\ and\
  \citenamefont {Holden}}]{Albeverio1988}%
  \BibitemOpen
  \bibfield  {author} {\bibinfo {author} {\bibfnamefont {S.}~\bibnamefont
  {Albeverio}}, \bibinfo {author} {\bibfnamefont {F.}~\bibnamefont {Gesztesy}},
  \bibinfo {author} {\bibfnamefont {R.}~\bibnamefont {H{\o}egh-Krohn}}, \ and\
  \bibinfo {author} {\bibfnamefont {H.}~\bibnamefont {Holden}},\ }\href@noop {}
  {\emph {\bibinfo {title} {Solvable Models in Quantum Mechanics}}}\ (\bibinfo
  {publisher} {Springer-Verlag, Berlin},\ \bibinfo {year} {1988})\BibitemShut
  {NoStop}%
\bibitem [{\citenamefont {Sakurai}\ and\ \citenamefont
  {Napolitano}(2011)}]{Sakurai2011}%
  \BibitemOpen
  \bibfield  {author} {\bibinfo {author} {\bibfnamefont {J.~J.}\ \bibnamefont
  {Sakurai}}\ and\ \bibinfo {author} {\bibfnamefont {J.}~\bibnamefont
  {Napolitano}},\ }\href@noop {} {\emph {\bibinfo {title} {Modern Quantum
  Mechanics}}}\ (\bibinfo  {publisher} {Addison-Wesley, Boston},\ \bibinfo
  {year} {2011})\BibitemShut {NoStop}%
\bibitem [{\citenamefont {Scully}\ and\ \citenamefont
  {Zubairy}(1997)}]{Scully1997}%
  \BibitemOpen
  \bibfield  {author} {\bibinfo {author} {\bibfnamefont {M.~O.}\ \bibnamefont
  {Scully}}\ and\ \bibinfo {author} {\bibfnamefont {M.~S.}\ \bibnamefont
  {Zubairy}},\ }\href@noop {} {\emph {\bibinfo {title} {Quantum Optics}}}\
  (\bibinfo  {publisher} {Cambridge University Press, Cambridge},\ \bibinfo
  {year} {1997})\BibitemShut {NoStop}%
\bibitem [{\citenamefont {Ashcroft}\ and\ \citenamefont
  {Mermin}(1976)}]{Ashcroft1976}%
  \BibitemOpen
  \bibfield  {author} {\bibinfo {author} {\bibfnamefont {N.~W.}\ \bibnamefont
  {Ashcroft}}\ and\ \bibinfo {author} {\bibfnamefont {N.~D.}\ \bibnamefont
  {Mermin}},\ }\href@noop {} {\emph {\bibinfo {title} {Solid State Physics}}}\
  (\bibinfo  {publisher} {Saunders, Philadelphia},\ \bibinfo {year}
  {1976})\BibitemShut {NoStop}%
\bibitem [{\citenamefont {Kato}(1966)}]{Kato1966}%
  \BibitemOpen
  \bibfield  {author} {\bibinfo {author} {\bibfnamefont {T.}~\bibnamefont
  {Kato}},\ }\href@noop {} {\emph {\bibinfo {title} {Perturbation Theory for
  Linear Operators}}}\ (\bibinfo  {publisher} {Springer},\ \bibinfo {address}
  {New York},\ \bibinfo {year} {1966})\BibitemShut {NoStop}%
\bibitem [{\citenamefont {Polkovnikov}\ \emph {et~al.}(2011)\citenamefont
  {Polkovnikov}, \citenamefont {Sengupta}, \citenamefont {Silva},\ and\
  \citenamefont {Vengalattore}}]{Polkovnikov2011}%
  \BibitemOpen
  \bibfield  {author} {\bibinfo {author} {\bibfnamefont {A.}~\bibnamefont
  {Polkovnikov}}, \bibinfo {author} {\bibfnamefont {K.}~\bibnamefont
  {Sengupta}}, \bibinfo {author} {\bibfnamefont {A.}~\bibnamefont {Silva}}, \
  and\ \bibinfo {author} {\bibfnamefont {M.}~\bibnamefont {Vengalattore}},\
  }\href {\doibase 10.1103/RevModPhys.83.863} {\bibfield  {journal} {\bibinfo
  {journal} {Rev. Mod. Phys.}\ }\textbf {\bibinfo {volume} {83}},\ \bibinfo
  {pages} {863} (\bibinfo {year} {2011})}\BibitemShut {NoStop}%
\bibitem [{\citenamefont {Jaksch}\ \emph {et~al.}(1998)\citenamefont {Jaksch},
  \citenamefont {Bruder}, \citenamefont {Cirac}, \citenamefont {Gardiner},\
  and\ \citenamefont {Zoller}}]{Jaksch1998}%
  \BibitemOpen
  \bibfield  {author} {\bibinfo {author} {\bibfnamefont {D.}~\bibnamefont
  {Jaksch}}, \bibinfo {author} {\bibfnamefont {C.}~\bibnamefont {Bruder}},
  \bibinfo {author} {\bibfnamefont {J.~I.}\ \bibnamefont {Cirac}}, \bibinfo
  {author} {\bibfnamefont {C.~W.}\ \bibnamefont {Gardiner}}, \ and\ \bibinfo
  {author} {\bibfnamefont {P.}~\bibnamefont {Zoller}},\ }\href {\doibase
  10.1103/PhysRevLett.81.3108} {\bibfield  {journal} {\bibinfo  {journal}
  {Phys. Rev. Lett.}\ }\textbf {\bibinfo {volume} {81}},\ \bibinfo {pages}
  {3108} (\bibinfo {year} {1998})}\BibitemShut {NoStop}%
\bibitem [{\citenamefont {Hofstetter}\ \emph {et~al.}(2002)\citenamefont
  {Hofstetter}, \citenamefont {Cirac}, \citenamefont {Zoller}, \citenamefont
  {Demler},\ and\ \citenamefont {Lukin}}]{Hofstetter2002}%
  \BibitemOpen
  \bibfield  {author} {\bibinfo {author} {\bibfnamefont {W.}~\bibnamefont
  {Hofstetter}}, \bibinfo {author} {\bibfnamefont {J.~I.}\ \bibnamefont
  {Cirac}}, \bibinfo {author} {\bibfnamefont {P.}~\bibnamefont {Zoller}},
  \bibinfo {author} {\bibfnamefont {E.}~\bibnamefont {Demler}}, \ and\ \bibinfo
  {author} {\bibfnamefont {M.~D.}\ \bibnamefont {Lukin}},\ }\href {\doibase
  10.1103/PhysRevLett.89.220407} {\bibfield  {journal} {\bibinfo  {journal}
  {Phys. Rev. Lett.}\ }\textbf {\bibinfo {volume} {89}},\ \bibinfo {pages}
  {220407} (\bibinfo {year} {2002})}\BibitemShut {NoStop}%
\bibitem [{\citenamefont {Cheneau}\ \emph {et~al.}(2012)\citenamefont
  {Cheneau}, \citenamefont {Barmettler}, \citenamefont {Poletti}, \citenamefont
  {Endres}, \citenamefont {Schau\ss}, \citenamefont {Fukuhara}, \citenamefont
  {Gross}, \citenamefont {Bloch}, \citenamefont {Kollath},\ and\ \citenamefont
  {Kuhr}}]{Bloch2012b}%
  \BibitemOpen
  \bibfield  {author} {\bibinfo {author} {\bibfnamefont {M.}~\bibnamefont
  {Cheneau}}, \bibinfo {author} {\bibfnamefont {P.}~\bibnamefont {Barmettler}},
  \bibinfo {author} {\bibfnamefont {D.}~\bibnamefont {Poletti}}, \bibinfo
  {author} {\bibfnamefont {M.}~\bibnamefont {Endres}}, \bibinfo {author}
  {\bibfnamefont {P.}~\bibnamefont {Schau\ss}}, \bibinfo {author}
  {\bibfnamefont {T.}~\bibnamefont {Fukuhara}}, \bibinfo {author}
  {\bibfnamefont {C.}~\bibnamefont {Gross}}, \bibinfo {author} {\bibfnamefont
  {I.}~\bibnamefont {Bloch}}, \bibinfo {author} {\bibfnamefont
  {C.}~\bibnamefont {Kollath}}, \ and\ \bibinfo {author} {\bibfnamefont
  {S.}~\bibnamefont {Kuhr}},\ }\href {\doibase 10.1038/nature10748} {\bibfield
  {journal} {\bibinfo  {journal} {Nature}\ }\textbf {\bibinfo {volume} {481}},\
  \bibinfo {pages} {484} (\bibinfo {year} {2012})}\BibitemShut {NoStop}%
\bibitem [{\citenamefont {Vijayan}\ \emph {et~al.}(2020)\citenamefont
  {Vijayan}, \citenamefont {Sompet}, \citenamefont {Salomon}, \citenamefont
  {Koepsell}, \citenamefont {Hirthe}, \citenamefont {Bohrdt}, \citenamefont
  {Grusdt}, \citenamefont {Bloch},\ and\ \citenamefont {Gross}}]{Gross2020}%
  \BibitemOpen
  \bibfield  {author} {\bibinfo {author} {\bibfnamefont {J.}~\bibnamefont
  {Vijayan}}, \bibinfo {author} {\bibfnamefont {P.}~\bibnamefont {Sompet}},
  \bibinfo {author} {\bibfnamefont {G.}~\bibnamefont {Salomon}}, \bibinfo
  {author} {\bibfnamefont {J.}~\bibnamefont {Koepsell}}, \bibinfo {author}
  {\bibfnamefont {S.}~\bibnamefont {Hirthe}}, \bibinfo {author} {\bibfnamefont
  {A.}~\bibnamefont {Bohrdt}}, \bibinfo {author} {\bibfnamefont
  {F.}~\bibnamefont {Grusdt}}, \bibinfo {author} {\bibfnamefont
  {I.}~\bibnamefont {Bloch}}, \ and\ \bibinfo {author} {\bibfnamefont
  {C.}~\bibnamefont {Gross}},\ }\href {\doibase 10.1126/science.aay2354}
  {\bibfield  {journal} {\bibinfo  {journal} {Science}\ }\textbf {\bibinfo
  {volume} {367}},\ \bibinfo {pages} {186} (\bibinfo {year}
  {2020})}\BibitemShut {NoStop}%
\bibitem [{\citenamefont {Bernien}\ \emph {et~al.}(2017)\citenamefont
  {Bernien}, \citenamefont {Schwartz}, \citenamefont {Keesling}, \citenamefont
  {Levine}, \citenamefont {Omran}, \citenamefont {Pichler}, \citenamefont
  {Choi}, \citenamefont {Zibrov}, \citenamefont {Endres}, \citenamefont
  {Greiner}, \citenamefont {Vuleti\'c},\ and\ \citenamefont
  {Lukin}}]{Bernien2017}%
  \BibitemOpen
  \bibfield  {author} {\bibinfo {author} {\bibfnamefont {H.}~\bibnamefont
  {Bernien}}, \bibinfo {author} {\bibfnamefont {S.}~\bibnamefont {Schwartz}},
  \bibinfo {author} {\bibfnamefont {A.}~\bibnamefont {Keesling}}, \bibinfo
  {author} {\bibfnamefont {H.}~\bibnamefont {Levine}}, \bibinfo {author}
  {\bibfnamefont {A.}~\bibnamefont {Omran}}, \bibinfo {author} {\bibfnamefont
  {H.}~\bibnamefont {Pichler}}, \bibinfo {author} {\bibfnamefont
  {S.}~\bibnamefont {Choi}}, \bibinfo {author} {\bibfnamefont {A.~S.}\
  \bibnamefont {Zibrov}}, \bibinfo {author} {\bibfnamefont {M.}~\bibnamefont
  {Endres}}, \bibinfo {author} {\bibfnamefont {M.}~\bibnamefont {Greiner}},
  \bibinfo {author} {\bibfnamefont {V.}~\bibnamefont {Vuleti\'c}}, \ and\
  \bibinfo {author} {\bibfnamefont {M.~D.}\ \bibnamefont {Lukin}},\ }\href
  {http://dx.doi.org/10.1038/nature24622} {\bibfield  {journal} {\bibinfo
  {journal} {Nature}\ }\textbf {\bibinfo {volume} {551}},\ \bibinfo {pages}
  {579} (\bibinfo {year} {2017})}\BibitemShut {NoStop}%
\bibitem [{\citenamefont {Turner}\ \emph {et~al.}(2018)\citenamefont {Turner},
  \citenamefont {Michailidis}, \citenamefont {Abanin}, \citenamefont {Serbyn},\
  and\ \citenamefont {Papi\'c}}]{Turner2018}%
  \BibitemOpen
  \bibfield  {author} {\bibinfo {author} {\bibfnamefont {C.~J.}\ \bibnamefont
  {Turner}}, \bibinfo {author} {\bibfnamefont {A.~A.}\ \bibnamefont
  {Michailidis}}, \bibinfo {author} {\bibfnamefont {D.~A.}\ \bibnamefont
  {Abanin}}, \bibinfo {author} {\bibfnamefont {M.}~\bibnamefont {Serbyn}}, \
  and\ \bibinfo {author} {\bibfnamefont {Z.}~\bibnamefont {Papi\'c}},\ }\href
  {\doibase 10.1038/s41567-018-0137-5} {\bibfield  {journal} {\bibinfo
  {journal} {Nat. Phys.}\ }\textbf {\bibinfo {volume} {14}},\ \bibinfo {pages}
  {745} (\bibinfo {year} {2018})}\BibitemShut {NoStop}%
\bibitem [{\citenamefont {Schrieffer}\ and\ \citenamefont
  {Wolff}(1966)}]{Schrieffer1966}%
  \BibitemOpen
  \bibfield  {author} {\bibinfo {author} {\bibfnamefont {J.~R.}\ \bibnamefont
  {Schrieffer}}\ and\ \bibinfo {author} {\bibfnamefont {P.~A.}\ \bibnamefont
  {Wolff}},\ }\href {\doibase 10.1103/PhysRev.149.491} {\bibfield  {journal}
  {\bibinfo  {journal} {Phys. Rev.}\ }\textbf {\bibinfo {volume} {149}},\
  \bibinfo {pages} {491} (\bibinfo {year} {1966})}\BibitemShut {NoStop}%
\bibitem [{\citenamefont {Bravyi}\ \emph {et~al.}(2011)\citenamefont {Bravyi},
  \citenamefont {DiVincenzo},\ and\ \citenamefont {Loss}}]{Bravyi2011}%
  \BibitemOpen
  \bibfield  {author} {\bibinfo {author} {\bibfnamefont {S.}~\bibnamefont
  {Bravyi}}, \bibinfo {author} {\bibfnamefont {D.~P.}\ \bibnamefont
  {DiVincenzo}}, \ and\ \bibinfo {author} {\bibfnamefont {D.}~\bibnamefont
  {Loss}},\ }\href {\doibase https://doi.org/10.1016/j.aop.2011.06.004}
  {\bibfield  {journal} {\bibinfo  {journal} {Ann. Phys.}\ }\textbf {\bibinfo
  {volume} {326}},\ \bibinfo {pages} {2793 } (\bibinfo {year}
  {2011})}\BibitemShut {NoStop}%
\bibitem [{\citenamefont {Lieb}\ and\ \citenamefont
  {Robinson}(1972)}]{Lieb1972}%
  \BibitemOpen
  \bibfield  {author} {\bibinfo {author} {\bibfnamefont {E.~H.}\ \bibnamefont
  {Lieb}}\ and\ \bibinfo {author} {\bibfnamefont {D.~W.}\ \bibnamefont
  {Robinson}},\ }\href {https://doi.org/10.1007/978-3-662-10018-9_25}
  {\bibfield  {journal} {\bibinfo  {journal} {Commun. Math. Phys.}\ }\textbf
  {\bibinfo {volume} {28}},\ \bibinfo {pages} {251} (\bibinfo {year}
  {1972})}\BibitemShut {NoStop}%
\bibitem [{\citenamefont {Nachtergaele}\ and\ \citenamefont
  {Sims}(2006)}]{Nachtergaele2006}%
  \BibitemOpen
  \bibfield  {author} {\bibinfo {author} {\bibfnamefont {B.}~\bibnamefont
  {Nachtergaele}}\ and\ \bibinfo {author} {\bibfnamefont {R.}~\bibnamefont
  {Sims}},\ }\href {\doibase 10.1007/s00220-006-1556-1} {\bibfield  {journal}
  {\bibinfo  {journal} {Commun. Math. Phys.}\ }\textbf {\bibinfo {volume}
  {265}},\ \bibinfo {pages} {119} (\bibinfo {year} {2006})}\BibitemShut
  {NoStop}%
\bibitem [{\citenamefont {Bravyi}\ \emph {et~al.}(2006)\citenamefont {Bravyi},
  \citenamefont {Hastings},\ and\ \citenamefont {Verstraete}}]{Bravyi2006}%
  \BibitemOpen
  \bibfield  {author} {\bibinfo {author} {\bibfnamefont {S.}~\bibnamefont
  {Bravyi}}, \bibinfo {author} {\bibfnamefont {M.~B.}\ \bibnamefont
  {Hastings}}, \ and\ \bibinfo {author} {\bibfnamefont {F.}~\bibnamefont
  {Verstraete}},\ }\href {\doibase 10.1103/PhysRevLett.97.050401} {\bibfield
  {journal} {\bibinfo  {journal} {Phys. Rev. Lett.}\ }\textbf {\bibinfo
  {volume} {97}},\ \bibinfo {pages} {050401} (\bibinfo {year}
  {2006})}\BibitemShut {NoStop}%
\bibitem [{\citenamefont {Margolus}\ and\ \citenamefont
  {Levitin}(1998)}]{Margolus1998}%
  \BibitemOpen
  \bibfield  {author} {\bibinfo {author} {\bibfnamefont {N.}~\bibnamefont
  {Margolus}}\ and\ \bibinfo {author} {\bibfnamefont {L.~B.}\ \bibnamefont
  {Levitin}},\ }\href {\doibase https://doi.org/10.1016/S0167-2789(98)00054-2}
  {\bibfield  {journal} {\bibinfo  {journal} {Physica D}\ }\textbf {\bibinfo
  {volume} {120}},\ \bibinfo {pages} {188 } (\bibinfo {year}
  {1998})}\BibitemShut {NoStop}%
\bibitem [{\citenamefont {Taddei}\ \emph {et~al.}(2013)\citenamefont {Taddei},
  \citenamefont {Escher}, \citenamefont {Davidovich},\ and\ \citenamefont
  {de~Matos~Filho}}]{Taddei2013}%
  \BibitemOpen
  \bibfield  {author} {\bibinfo {author} {\bibfnamefont {M.~M.}\ \bibnamefont
  {Taddei}}, \bibinfo {author} {\bibfnamefont {B.~M.}\ \bibnamefont {Escher}},
  \bibinfo {author} {\bibfnamefont {L.}~\bibnamefont {Davidovich}}, \ and\
  \bibinfo {author} {\bibfnamefont {R.~L.}\ \bibnamefont {de~Matos~Filho}},\
  }\href {\doibase 10.1103/PhysRevLett.110.050402} {\bibfield  {journal}
  {\bibinfo  {journal} {Phys. Rev. Lett.}\ }\textbf {\bibinfo {volume} {110}},\
  \bibinfo {pages} {050402} (\bibinfo {year} {2013})}\BibitemShut {NoStop}%
\bibitem [{\citenamefont {del Campo}\ \emph {et~al.}(2013)\citenamefont {del
  Campo}, \citenamefont {Egusquiza}, \citenamefont {Plenio},\ and\
  \citenamefont {Huelga}}]{delCampo2013}%
  \BibitemOpen
  \bibfield  {author} {\bibinfo {author} {\bibfnamefont {A.}~\bibnamefont {del
  Campo}}, \bibinfo {author} {\bibfnamefont {I.~L.}\ \bibnamefont {Egusquiza}},
  \bibinfo {author} {\bibfnamefont {M.~B.}\ \bibnamefont {Plenio}}, \ and\
  \bibinfo {author} {\bibfnamefont {S.~F.}\ \bibnamefont {Huelga}},\ }\href
  {\doibase 10.1103/PhysRevLett.110.050403} {\bibfield  {journal} {\bibinfo
  {journal} {Phys. Rev. Lett.}\ }\textbf {\bibinfo {volume} {110}},\ \bibinfo
  {pages} {050403} (\bibinfo {year} {2013})}\BibitemShut {NoStop}%
\bibitem [{\citenamefont {Deffner}\ and\ \citenamefont
  {Lutz}(2013)}]{Deffner2013}%
  \BibitemOpen
  \bibfield  {author} {\bibinfo {author} {\bibfnamefont {S.}~\bibnamefont
  {Deffner}}\ and\ \bibinfo {author} {\bibfnamefont {E.}~\bibnamefont {Lutz}},\
  }\href {\doibase 10.1103/PhysRevLett.111.010402} {\bibfield  {journal}
  {\bibinfo  {journal} {Phys. Rev. Lett.}\ }\textbf {\bibinfo {volume} {111}},\
  \bibinfo {pages} {010402} (\bibinfo {year} {2013})}\BibitemShut {NoStop}%
\bibitem [{\citenamefont {Abanin}\ \emph {et~al.}(2015)\citenamefont {Abanin},
  \citenamefont {De~Roeck},\ and\ \citenamefont {Huveneers}}]{Abanin2015b}%
  \BibitemOpen
  \bibfield  {author} {\bibinfo {author} {\bibfnamefont {D.~A.}\ \bibnamefont
  {Abanin}}, \bibinfo {author} {\bibfnamefont {W.}~\bibnamefont {De~Roeck}}, \
  and\ \bibinfo {author} {\bibfnamefont {F.}~\bibnamefont {Huveneers}},\ }\href
  {\doibase 10.1103/PhysRevLett.115.256803} {\bibfield  {journal} {\bibinfo
  {journal} {Phys. Rev. Lett.}\ }\textbf {\bibinfo {volume} {115}},\ \bibinfo
  {pages} {256803} (\bibinfo {year} {2015})}\BibitemShut {NoStop}%
\bibitem [{\citenamefont {Mori}\ \emph {et~al.}(2016)\citenamefont {Mori},
  \citenamefont {Kuwahara},\ and\ \citenamefont {Saito}}]{Mori2016}%
  \BibitemOpen
  \bibfield  {author} {\bibinfo {author} {\bibfnamefont {T.}~\bibnamefont
  {Mori}}, \bibinfo {author} {\bibfnamefont {T.}~\bibnamefont {Kuwahara}}, \
  and\ \bibinfo {author} {\bibfnamefont {K.}~\bibnamefont {Saito}},\ }\href
  {\doibase 10.1103/PhysRevLett.116.120401} {\bibfield  {journal} {\bibinfo
  {journal} {Phys. Rev. Lett.}\ }\textbf {\bibinfo {volume} {116}},\ \bibinfo
  {pages} {120401} (\bibinfo {year} {2016})}\BibitemShut {NoStop}%
\bibitem [{\citenamefont {Maldacena}\ \emph {et~al.}(2016)\citenamefont
  {Maldacena}, \citenamefont {Shenker},\ and\ \citenamefont
  {Stanford}}]{Stanford2016}%
  \BibitemOpen
  \bibfield  {author} {\bibinfo {author} {\bibfnamefont {J.}~\bibnamefont
  {Maldacena}}, \bibinfo {author} {\bibfnamefont {S.~H.}\ \bibnamefont
  {Shenker}}, \ and\ \bibinfo {author} {\bibfnamefont {D.}~\bibnamefont
  {Stanford}},\ }\href {\doibase 10.1007/JHEP08(2016)106} {\bibfield  {journal}
  {\bibinfo  {journal} {J. High Energy Phys.}\ }\textbf {\bibinfo {volume}
  {2016}},\ \bibinfo {pages} {106} (\bibinfo {year} {2016})}\BibitemShut
  {NoStop}%
\bibitem [{\citenamefont {Szehr}\ \emph {et~al.}(2015)\citenamefont {Szehr},
  \citenamefont {Reeb},\ and\ \citenamefont {Wolf}}]{Wolf2015}%
  \BibitemOpen
  \bibfield  {author} {\bibinfo {author} {\bibfnamefont {O.}~\bibnamefont
  {Szehr}}, \bibinfo {author} {\bibfnamefont {D.}~\bibnamefont {Reeb}}, \ and\
  \bibinfo {author} {\bibfnamefont {M.~M.}\ \bibnamefont {Wolf}},\ }\href
  {https://doi.org/10.1007/s00220-014-2188-5} {\bibfield  {journal} {\bibinfo
  {journal} {Commun. Math. Phys.}\ }\textbf {\bibinfo {volume} {333}},\
  \bibinfo {pages} {565} (\bibinfo {year} {2015})}\BibitemShut {NoStop}%
\bibitem [{\citenamefont {Gong}\ \emph {et~al.}(2020)\citenamefont {Gong},
  \citenamefont {Yoshioka}, \citenamefont {Shibata},\ and\ \citenamefont
  {Hamazaki}}]{Gong2020}%
  \BibitemOpen
  \bibfield  {author} {\bibinfo {author} {\bibfnamefont {Z.}~\bibnamefont
  {Gong}}, \bibinfo {author} {\bibfnamefont {N.}~\bibnamefont {Yoshioka}},
  \bibinfo {author} {\bibfnamefont {N.}~\bibnamefont {Shibata}}, \ and\
  \bibinfo {author} {\bibfnamefont {R.}~\bibnamefont {Hamazaki}},\ }\href@noop
  {} {\enquote {\bibinfo {title} {Error bounds for constrained dynamics in
  gapped quantum systems: Rigorous results and generalizations},}\ } (\bibinfo
  {year} {2020}),\ \bibinfo {note} {arXiv:2001.03421}\BibitemShut {NoStop}%
\bibitem [{\citenamefont {Feshbach}(1958)}]{Feshbach1958}%
  \BibitemOpen
  \bibfield  {author} {\bibinfo {author} {\bibfnamefont {H.}~\bibnamefont
  {Feshbach}},\ }\href {\doibase https://doi.org/10.1016/0003-4916(58)90007-1}
  {\bibfield  {journal} {\bibinfo  {journal} {Ann. Phys.}\ }\textbf {\bibinfo
  {volume} {5}},\ \bibinfo {pages} {357 } (\bibinfo {year} {1958})}\BibitemShut
  {NoStop}%
\bibitem [{\citenamefont {Feshbach}(1962)}]{Feshbach1962}%
  \BibitemOpen
  \bibfield  {author} {\bibinfo {author} {\bibfnamefont {H.}~\bibnamefont
  {Feshbach}},\ }\href {\doibase https://doi.org/10.1016/0003-4916(62)90221-X}
  {\bibfield  {journal} {\bibinfo  {journal} {Ann. Phys.}\ }\textbf {\bibinfo
  {volume} {19}},\ \bibinfo {pages} {287 } (\bibinfo {year}
  {1962})}\BibitemShut {NoStop}%
\bibitem [{\citenamefont {Brion}\ \emph {et~al.}(2007)\citenamefont {Brion},
  \citenamefont {Pedersen},\ and\ \citenamefont {M{\o}lmer}}]{Brion2007}%
  \BibitemOpen
  \bibfield  {author} {\bibinfo {author} {\bibfnamefont {E.}~\bibnamefont
  {Brion}}, \bibinfo {author} {\bibfnamefont {L.~H.}\ \bibnamefont {Pedersen}},
  \ and\ \bibinfo {author} {\bibfnamefont {K.}~\bibnamefont {M{\o}lmer}},\
  }\href {\doibase 10.1088/1751-8113/40/5/011} {\bibfield  {journal} {\bibinfo
  {journal} {J. Phys. A}\ }\textbf {\bibinfo {volume} {40}},\ \bibinfo {pages}
  {1033} (\bibinfo {year} {2007})}\BibitemShut {NoStop}%
\bibitem [{RF()}]{RF}%
  \BibitemOpen
  \href@noop {} {}\bibinfo {note} {Here we are working within the rotating
  frame, so the Hamiltonian is time-independent and involves the detuning
  rather than the atomic resonant frequency.}\BibitemShut {Stop}%
\bibitem [{\citenamefont {Sylvester}(1884)}]{Sylvester1884}%
  \BibitemOpen
  \bibfield  {author} {\bibinfo {author} {\bibfnamefont {J.}~\bibnamefont
  {Sylvester}},\ }\href@noop {} {\bibfield  {journal} {\bibinfo  {journal} {C.
  R. Acad. Sci. Paris}\ }\textbf {\bibinfo {volume} {99}},\ \bibinfo {pages}
  {67} (\bibinfo {year} {1884})}\BibitemShut {NoStop}%
\bibitem [{\citenamefont {Bhatia}(1997)}]{Bhatia1997}%
  \BibitemOpen
  \bibfield  {author} {\bibinfo {author} {\bibfnamefont {R.}~\bibnamefont
  {Bhatia}},\ }\href@noop {} {\emph {\bibinfo {title} {Matrix Analysis}}}\
  (\bibinfo  {publisher} {Springer, New York},\ \bibinfo {year}
  {1997})\BibitemShut {NoStop}%
\bibitem [{\citenamefont {Weyl}(1912)}]{Weyl1912}%
  \BibitemOpen
  \bibfield  {author} {\bibinfo {author} {\bibfnamefont {H.}~\bibnamefont
  {Weyl}},\ }\href {\doibase 10.1007/BF01456804} {\bibfield  {journal}
  {\bibinfo  {journal} {Math. Ann.}\ }\textbf {\bibinfo {volume} {71}},\
  \bibinfo {pages} {441} (\bibinfo {year} {1912})}\BibitemShut {NoStop}%
\bibitem [{\citenamefont {Gorin}\ \emph {et~al.}(2006)\citenamefont {Gorin},
  \citenamefont {Prosen}, \citenamefont {Seligman},\ and\ \citenamefont
  {\v{Z}nidari\v{c}}}]{Prosen2006}%
  \BibitemOpen
  \bibfield  {author} {\bibinfo {author} {\bibfnamefont {T.}~\bibnamefont
  {Gorin}}, \bibinfo {author} {\bibfnamefont {T.}~\bibnamefont {Prosen}},
  \bibinfo {author} {\bibfnamefont {T.~H.}\ \bibnamefont {Seligman}}, \ and\
  \bibinfo {author} {\bibfnamefont {M.}~\bibnamefont {\v{Z}nidari\v{c}}},\
  }\href {\doibase https://doi.org/10.1016/j.physrep.2006.09.003} {\bibfield
  {journal} {\bibinfo  {journal} {Phys. Rep.}\ }\textbf {\bibinfo {volume}
  {435}},\ \bibinfo {pages} {33 } (\bibinfo {year} {2006})}\BibitemShut
  {NoStop}%
\bibitem [{dLt()}]{dLt}%
  \BibitemOpen
  \href@noop {} {}\bibinfo {note} {Taking the time derivative for
  $L(t)=e^{-iH_1t}e^{i(H_1+V')t}$, we obtain
  $i\frac{d}{dt}L(t)=-e^{-iH_1t}V'e^{i(H_1+V')t}=-e^{-iH_1t}V'e^{iH_1t}L(t)$,
  implying that $L(t)$ is generated by $-e^{-iH_1t}V'e^{iH_1t}$.}\BibitemShut
  {Stop}%
\bibitem [{\citenamefont {Datta}\ \emph {et~al.}(1996)\citenamefont {Datta},
  \citenamefont {Fr\"ohlich}, \citenamefont {Rey-Bellet},\ and\ \citenamefont
  {Fern{\'a}ndez}}]{Datta1996}%
  \BibitemOpen
  \bibfield  {author} {\bibinfo {author} {\bibfnamefont {N.}~\bibnamefont
  {Datta}}, \bibinfo {author} {\bibfnamefont {J.}~\bibnamefont {Fr\"ohlich}},
  \bibinfo {author} {\bibfnamefont {L.}~\bibnamefont {Rey-Bellet}}, \ and\
  \bibinfo {author} {\bibfnamefont {R.}~\bibnamefont {Fern{\'a}ndez}},\
  }\href@noop {} {\bibfield  {journal} {\bibinfo  {journal} {Helv. Phys. Acta}\
  }\textbf {\bibinfo {volume} {69}},\ \bibinfo {pages} {752} (\bibinfo {year}
  {1996})}\BibitemShut {NoStop}%
\bibitem [{esa()}]{esal}%
  \BibitemOpen
  \href@noop {} {}\bibinfo {note} {By ``essentially" and ``almost", we mean
  that the Lieb-Robinson velocity $v$ may still weakly depend on $H_0$. See
  Sec. III B in Ref.~\cite{Gong2020}, especially Figs.~5 and 6, for further
  details.}\BibitemShut {Stop}%
\bibitem [{\citenamefont {Zanardi}\ and\ \citenamefont
  {Rasetti}(1997)}]{Zanardi1997}%
  \BibitemOpen
  \bibfield  {author} {\bibinfo {author} {\bibfnamefont {P.}~\bibnamefont
  {Zanardi}}\ and\ \bibinfo {author} {\bibfnamefont {M.}~\bibnamefont
  {Rasetti}},\ }\href {\doibase 10.1103/PhysRevLett.79.3306} {\bibfield
  {journal} {\bibinfo  {journal} {Phys. Rev. Lett.}\ }\textbf {\bibinfo
  {volume} {79}},\ \bibinfo {pages} {3306} (\bibinfo {year}
  {1997})}\BibitemShut {NoStop}%
\bibitem [{\citenamefont {Lidar}\ \emph {et~al.}(1998)\citenamefont {Lidar},
  \citenamefont {Chuang},\ and\ \citenamefont {Whaley}}]{Lidar1998}%
  \BibitemOpen
  \bibfield  {author} {\bibinfo {author} {\bibfnamefont {D.~A.}\ \bibnamefont
  {Lidar}}, \bibinfo {author} {\bibfnamefont {I.~L.}\ \bibnamefont {Chuang}}, \
  and\ \bibinfo {author} {\bibfnamefont {K.~B.}\ \bibnamefont {Whaley}},\
  }\href {\doibase 10.1103/PhysRevLett.81.2594} {\bibfield  {journal} {\bibinfo
   {journal} {Phys. Rev. Lett.}\ }\textbf {\bibinfo {volume} {81}},\ \bibinfo
  {pages} {2594} (\bibinfo {year} {1998})}\BibitemShut {NoStop}%
\bibitem [{\citenamefont {Misra}\ and\ \citenamefont
  {Sudarshan}(1977)}]{Misra1977}%
  \BibitemOpen
  \bibfield  {author} {\bibinfo {author} {\bibfnamefont {B.}~\bibnamefont
  {Misra}}\ and\ \bibinfo {author} {\bibfnamefont {E.~C.~G.}\ \bibnamefont
  {Sudarshan}},\ }\href {\doibase 10.1063/1.523304} {\bibfield  {journal}
  {\bibinfo  {journal} {J. Math. Phys.}\ }\textbf {\bibinfo {volume} {18}},\
  \bibinfo {pages} {756} (\bibinfo {year} {1977})}\BibitemShut {NoStop}%
\bibitem [{\citenamefont {Beige}\ \emph {et~al.}(2000)\citenamefont {Beige},
  \citenamefont {Braun}, \citenamefont {Tregenna},\ and\ \citenamefont
  {Knight}}]{Knight2000}%
  \BibitemOpen
  \bibfield  {author} {\bibinfo {author} {\bibfnamefont {A.}~\bibnamefont
  {Beige}}, \bibinfo {author} {\bibfnamefont {D.}~\bibnamefont {Braun}},
  \bibinfo {author} {\bibfnamefont {B.}~\bibnamefont {Tregenna}}, \ and\
  \bibinfo {author} {\bibfnamefont {P.~L.}\ \bibnamefont {Knight}},\ }\href
  {\doibase 10.1103/PhysRevLett.85.1762} {\bibfield  {journal} {\bibinfo
  {journal} {Phys. Rev. Lett.}\ }\textbf {\bibinfo {volume} {85}},\ \bibinfo
  {pages} {1762} (\bibinfo {year} {2000})}\BibitemShut {NoStop}%
\bibitem [{\citenamefont {Facchi}\ and\ \citenamefont
  {Pascazio}(2002)}]{Facchi2002}%
  \BibitemOpen
  \bibfield  {author} {\bibinfo {author} {\bibfnamefont {P.}~\bibnamefont
  {Facchi}}\ and\ \bibinfo {author} {\bibfnamefont {S.}~\bibnamefont
  {Pascazio}},\ }\href {\doibase 10.1103/PhysRevLett.89.080401} {\bibfield
  {journal} {\bibinfo  {journal} {Phys. Rev. Lett.}\ }\textbf {\bibinfo
  {volume} {89}},\ \bibinfo {pages} {080401} (\bibinfo {year}
  {2002})}\BibitemShut {NoStop}%
\bibitem [{\citenamefont {Stannigel}\ \emph {et~al.}(2014)\citenamefont
  {Stannigel}, \citenamefont {Hauke}, \citenamefont {Marcos}, \citenamefont
  {Hafezi}, \citenamefont {Diehl}, \citenamefont {Dalmonte},\ and\
  \citenamefont {Zoller}}]{Zoller2014}%
  \BibitemOpen
  \bibfield  {author} {\bibinfo {author} {\bibfnamefont {K.}~\bibnamefont
  {Stannigel}}, \bibinfo {author} {\bibfnamefont {P.}~\bibnamefont {Hauke}},
  \bibinfo {author} {\bibfnamefont {D.}~\bibnamefont {Marcos}}, \bibinfo
  {author} {\bibfnamefont {M.}~\bibnamefont {Hafezi}}, \bibinfo {author}
  {\bibfnamefont {S.}~\bibnamefont {Diehl}}, \bibinfo {author} {\bibfnamefont
  {M.}~\bibnamefont {Dalmonte}}, \ and\ \bibinfo {author} {\bibfnamefont
  {P.}~\bibnamefont {Zoller}},\ }\href {\doibase
  10.1103/PhysRevLett.112.120406} {\bibfield  {journal} {\bibinfo  {journal}
  {Phys. Rev. Lett.}\ }\textbf {\bibinfo {volume} {112}},\ \bibinfo {pages}
  {120406} (\bibinfo {year} {2014})}\BibitemShut {NoStop}%
\bibitem [{\citenamefont {Gong}\ \emph {et~al.}(2017)\citenamefont {Gong},
  \citenamefont {Higashikawa},\ and\ \citenamefont {Ueda}}]{Gong2017}%
  \BibitemOpen
  \bibfield  {author} {\bibinfo {author} {\bibfnamefont {Z.}~\bibnamefont
  {Gong}}, \bibinfo {author} {\bibfnamefont {S.}~\bibnamefont {Higashikawa}}, \
  and\ \bibinfo {author} {\bibfnamefont {M.}~\bibnamefont {Ueda}},\ }\href
  {\doibase 10.1103/PhysRevLett.118.200401} {\bibfield  {journal} {\bibinfo
  {journal} {Phys. Rev. Lett.}\ }\textbf {\bibinfo {volume} {118}},\ \bibinfo
  {pages} {200401} (\bibinfo {year} {2017})}\BibitemShut {NoStop}%
\bibitem [{\citenamefont {Hauke}\ and\ \citenamefont
  {Tagliacozzo}(2013)}]{Hauke2013}%
  \BibitemOpen
  \bibfield  {author} {\bibinfo {author} {\bibfnamefont {P.}~\bibnamefont
  {Hauke}}\ and\ \bibinfo {author} {\bibfnamefont {L.}~\bibnamefont
  {Tagliacozzo}},\ }\href {\doibase 10.1103/PhysRevLett.111.207202} {\bibfield
  {journal} {\bibinfo  {journal} {Phys. Rev. Lett.}\ }\textbf {\bibinfo
  {volume} {111}},\ \bibinfo {pages} {207202} (\bibinfo {year}
  {2013})}\BibitemShut {NoStop}%
\bibitem [{\citenamefont {Eisert}\ \emph {et~al.}(2013)\citenamefont {Eisert},
  \citenamefont {van~den Worm}, \citenamefont {Manmana},\ and\ \citenamefont
  {Kastner}}]{Eisert2013}%
  \BibitemOpen
  \bibfield  {author} {\bibinfo {author} {\bibfnamefont {J.}~\bibnamefont
  {Eisert}}, \bibinfo {author} {\bibfnamefont {M.}~\bibnamefont {van~den
  Worm}}, \bibinfo {author} {\bibfnamefont {S.~R.}\ \bibnamefont {Manmana}}, \
  and\ \bibinfo {author} {\bibfnamefont {M.}~\bibnamefont {Kastner}},\ }\href
  {\doibase 10.1103/PhysRevLett.111.260401} {\bibfield  {journal} {\bibinfo
  {journal} {Phys. Rev. Lett.}\ }\textbf {\bibinfo {volume} {111}},\ \bibinfo
  {pages} {260401} (\bibinfo {year} {2013})}\BibitemShut {NoStop}%
\bibitem [{\citenamefont {Gong}\ \emph {et~al.}(2014)\citenamefont {Gong},
  \citenamefont {Foss-Feig}, \citenamefont {Michalakis},\ and\ \citenamefont
  {Gorshkov}}]{Gorshkov2014b}%
  \BibitemOpen
  \bibfield  {author} {\bibinfo {author} {\bibfnamefont {Z.-X.}\ \bibnamefont
  {Gong}}, \bibinfo {author} {\bibfnamefont {M.}~\bibnamefont {Foss-Feig}},
  \bibinfo {author} {\bibfnamefont {S.}~\bibnamefont {Michalakis}}, \ and\
  \bibinfo {author} {\bibfnamefont {A.~V.}\ \bibnamefont {Gorshkov}},\ }\href
  {\doibase 10.1103/PhysRevLett.113.030602} {\bibfield  {journal} {\bibinfo
  {journal} {Phys. Rev. Lett.}\ }\textbf {\bibinfo {volume} {113}},\ \bibinfo
  {pages} {030602} (\bibinfo {year} {2014})}\BibitemShut {NoStop}%
\bibitem [{\citenamefont {Chen}\ and\ \citenamefont {Lucas}(2019)}]{Lucas2019}%
  \BibitemOpen
  \bibfield  {author} {\bibinfo {author} {\bibfnamefont {C.-F.}\ \bibnamefont
  {Chen}}\ and\ \bibinfo {author} {\bibfnamefont {A.}~\bibnamefont {Lucas}},\
  }\href {\doibase 10.1103/PhysRevLett.123.250605} {\bibfield  {journal}
  {\bibinfo  {journal} {Phys. Rev. Lett.}\ }\textbf {\bibinfo {volume} {123}},\
  \bibinfo {pages} {250605} (\bibinfo {year} {2019})}\BibitemShut {NoStop}%
\bibitem [{\citenamefont {Kuwahara}\ and\ \citenamefont
  {Saito}(2019)}]{Kuwahara2019}%
  \BibitemOpen
  \bibfield  {author} {\bibinfo {author} {\bibfnamefont {T.}~\bibnamefont
  {Kuwahara}}\ and\ \bibinfo {author} {\bibfnamefont {K.}~\bibnamefont
  {Saito}},\ }\href@noop {} {\enquote {\bibinfo {title} {Strictly linear light
  cones in long-range interacting systems of arbitrary dimensions},}\ }
  (\bibinfo {year} {2019}),\ \bibinfo {note} {arXiv:1910.14477}\BibitemShut
  {NoStop}%
\bibitem [{\citenamefont {Johansson}\ \emph {et~al.}(2013)\citenamefont
  {Johansson}, \citenamefont {Nation},\ and\ \citenamefont {Nori}}]{qutip}%
  \BibitemOpen
  \bibfield  {author} {\bibinfo {author} {\bibfnamefont {J.~R.}\ \bibnamefont
  {Johansson}}, \bibinfo {author} {\bibfnamefont {P.~D.}\ \bibnamefont
  {Nation}}, \ and\ \bibinfo {author} {\bibfnamefont {F.}~\bibnamefont
  {Nori}},\ }\href {\doibase https://doi.org/10.1016/j.cpc.2012.11.019}
  {\bibfield  {journal} {\bibinfo  {journal} {Comput. Phys. Commun.}\ }\textbf
  {\bibinfo {volume} {184}},\ \bibinfo {pages} {1234 } (\bibinfo {year}
  {2013})}\BibitemShut {NoStop}%
\end{thebibliography}%

\end{document}